\documentclass[12pt]{article}
\title{The role of AGK theorem in QCD} 
\author{V.A. Abramovsky\thanks{email: victor.abramovsky@novsu.ru}, Alexey V. Popov\thanks{email: avp@novgorod.net} \\ \\ 
\small{\emph{Velikiy Novgorod, Russia}} 
}
\date{}

\begin{document}
\maketitle
\abstract{We rise the question about a role of AGK theorem in QCD. 
Considering multiple-gluons emissions in a scattering process, we formulate the AGK-like rules of calculation  
of exclusive gluon distributions. 
The result shows that the naive extrapolation of the soft AGK, based on the pomeron idea, is not constructive 
and one should find a more generalized formulation.
}
\newcommand{\ket}[1]{\ensuremath{|#1\rangle}}
\newcommand{\bra}[1]{\ensuremath{\langle#1|}}
\newcommand{\braket}[2]{\ensuremath{\langle#1|#2\rangle}}
\section{Introduction}
AGK cutting rules have the long story \cite{AGK}.
After the first discovery, these rules became a very important part of high energy phenomenology.
The main idea is that a scattering process in a multi-constituent environment can be considered as 
a set of parallel instantaneous interactions (pomeron exchanges). 
The key elements here are ``pomeron'' and that a scattering amplitudes can be decomposed in terms of elementary pomerons.
Usually, these elements are formulated as a Reggeon Field Theory (RFT) in which a diagram technique and AGK rules generates all relevant amplitudes. 
However, an arbitrary RFT has a lot number of coupling constants (in fact, all they are functions). 
This means that one should associate new coupling for every multipomeron vertex and especially for every particle-pomerons branching vertex.   
We just believe that RFT is a fairly good effective theory for high-energy soft hardron phenomenology
and all necessary coupling constants can be extracted from the experimental data and used in subsequent calculations. 
The one more rule here is that the pomeron inelastic state is a 
multiperipheral state where produced secondary particles uniformly populate midrapidity region.  
In principle, in this effective theory we can only postulate AGK rules.
In other words, we just select RTF and AGK for an appropriate description of the experimental data.

From theoretical viewpoint, any RFT should by obtained from high energy limit of a fundamental quantum field theory.
However, the rigorous proofs of the RFT existence are known only for the toy models such as: the $\varphi^3$ scalar field theory, a dual string-like theory, and so on.
Similarly, all known proofs of AGK rules are also known for the toy models.
So, it is natural to unify proofs of RFT and AGK in the single question about high energy behavior of a quantum field theory. 
It should be stressed that the crucial building block in the formulations of RFT and AGK is a pomeron, 
since the original cutting rules are applicable only for an explicit n-pomeron exchange.
Hence, concerning QCD, we have to find a diagram technique and an object that will play the role of the pomeron.

A more intuitive formulation of high energy scattering can be found in the parton model.
A parton is just an elementary excitation of the considered effective field theory.
A high energy hardron is a compound object builded from partons. 
If we boost a hardron to high rapidity $Y$, then in its wave function there will be new partons that are strongly correlated with primary partons.
If the elementary scattering is a process involving a limited number of partons, then
it can be shown \cite{avp_agk} that in the parton model the AGK rules arise naturally due to the natural action of secondary-quantized scattering operator $\hat S$ on multiparton states.

The great efforts were made to investigate the high energy behavior of QCD.
The first classical results here are gluon reggeization and BFKL pomeron \cite{BFKL}. 
To obtain RFT from QCD, one should construct a diagram technique that uses pomerons as elementary building blocks.
However, this is very difficult and unsolved task, only primitive diagrams can be constructed by hand.
Instead, it is possible to construct a theory of interacting reggeized gluons \cite{Lipatov97}.
For example, BFKL pomeron is considered as a bound state of two reggeized gluons.
The problem here is that to extract AGK from this RFT we must construct the theory in the term of pomerons, not in the terms of reggeized gluons.
Unfortunately, even the simple BFKL pomeron is an infinity set of classical regge poles.
Moreover, there exists the intensive color duplication of poles \cite{avp_rz,Kovner0512} 
and vertexes of pomeron interaction can connect an arbitrary number of lines.
All this gives a large combinatoric growing of complexity in the hypothetical diagram technique. 
So, we can conclude that this way is too complicated for our task.
Here we only review known results concerning AGK cancellations in the BFKL-QCD approach.
In Ref. \cite{Braun} the reggeized-gluon technique was used to check AGK rules in a gluon emission from the triple pomeron vertex.
It was shown that AGK is valid only if the vertex is a fully symmetric in four outgoing reggeized gluons.
The similar result was found in Ref. \cite{Bartels08} where it was argued that a gluon emission from the vertex requires a modification of AGK rules.  
In Ref. \cite{Levin} this result had been partially confirmed in the dipole approach.
In Ref. \cite{Nikolaev} it was shown that the complicated non-Abelian color structure of diagrams gives two possible types of pomeron cuts. 
One can cut the two-pomeron exchange, considered as a four reggeized gluon state, in the two distinct way: through the pomeron or not. 
Also, the vertex contribution depends on the type of the cut.
Therefore, AGK rules should be modified. 
However, it was noted in Ref. \cite{Bartels05} that the such modification is already possible in the original formulation of AGK \cite{AGK} and
corresponds to an other regge pole such as odderon.
So, we see that a complete solution of the problem requires the much more information about the reggeons and the vertexes.

In the other more pragmatic approach, QCD as a microscopic theory of gluons allows to calculate directly amplitudes of gluon emissions without a construction of intermediate RTF.
This means that there is no sense to find a something like AGK theorem, since there are no pomerons, no pomeron loops, no vertexes, no cuttings, and no diagrams. 
The inclusive spectra of multiple-gluons emissions and quasielastic scattering amplitudes can be calculated from the standard quantum mechanic of field theory
in both the path integral and Hamiltonian approaches.  
Maybe, it is more natural to forget complectly about AGK in QCD, since there are no practical reasons to search it?
We think that in QCD it is sufficient to construct a systematic calculation procedure   
of multiple-gluons emissions in a high-energy scattering process.  

Nevertheless, is this paper we propose a reformulation of the AGK rules for high energy evolution in QCD.
The first idea is to generalize AGK to a more weak statement that
relates inclusive and exclusive parton distributions.
Indeed, recall that the original AGK rules give a cutting algorithm to obtain exactly $n$-number of pomeron cuts in a final state.
In the soft hardron phenomenology, each such pomeron cut is associated with an observable multiperipheral parton shower.
In QCD we expect to obtain a rule to calculate exclusive gluon distributions,  i.e. the probability to find exactly $n$ gluons in a final state. 
In additional, in the original AGK case we have the AGK cancellation that relates the inclusive spectrum and the elastic amplitude. 
Hence, second our idea is to divide AGK rules by the two parts: the relation between inclusive and exclusive spectrums and
the relation between inclusive spectrum and elastic amplitude.
The first part is combinatorial, the second part is dynamical and depends on a specific theory.  

In our approach we do not assume the AGK cancellation 
because in QCD there is no explicit relation between inclusive distributions and an elastic amplitude. 
Therefore, we reconstruct only one half of original AGK. 
Since in our method there are no pomerons, we just propose a tool to calculate the exclusive gluon spectrum from the inclusive one
in the context of high energy QCD.
Though this result is not the full original AGK, it well reproduces the famous AGK algebra of contributions to the total cross section.

This paper is organized as follows. 
In Sec. \ref{sec1}, using the formalism of wave function, we construct the general picture of multi-gluon emission in high energy scattering.
The main part on this picture is the boost operator $\Omega$ that creates additional gluons in lightcone wave function of a hardron.
In Sec. \ref{sec2} we explicitly calculate the boost operator $\Omega$ from the first order of lightcone QCD Hamiltonian.
In Sec. \ref{sec3} we show how the operator $\Omega$ reproduces eikonal $S$-matrix in QCD.
In Sec. \ref{sec4} we derive AGK-like relation between inclusive and exclusive gluon distributions.
Section \ref{sec5} contains our conclusions.

\section{Scattering setup} \label{sec1}
Consider a standard fast projective \ket {P_0} considered as a set of QCD partons. Under a rapidity boost $Y$ we have
\begin{equation}
 \ket P = \Omega \ket {P_0}
\end{equation}
where $\Omega (Y)$ is the unitary boost operator which creates new gluons at intermediate rapidity.

Scattering on a target state gives a new state
\begin{equation}
\ket {P'}= S \ket P = S \Omega \ket {P_0} \label{eq_31}
\end{equation}
where $S$ is the operator of quasielastic eikonal scattering. 
Corresponding elastic amplitude is obviously given by
\begin{equation} \label{eq_9}
S_{el}=\braket{P}{P'}=\bra {P_0} \Omega^\dag S \Omega \ket {P_0}
\end{equation}
where the final result should be averaged over the target fields, but this is unimportant at current stage.

Note, that the operator $S$ in Eq. (\ref{eq_31}) also acts on the new soft gluons created by the boost operator $\Omega$.
These soft gluons should be considered as a part of the boosted projectile.
We will clarify this note below in the end of Sec. \ref{sec3}.

Besides the elastic amplitude there are an other class of processes where a gluons emission takes place.
There are two classes of possible observables: inclusive and exclusive. Both can be related to each other.
Inclusive observables is constructed from the operator of gluon number $N$
\begin{equation}
N(Y,\vec x) = \sum_{i,a} a_{Y,x}^\dag a_{Y,x}^{\phantom\dag}
\end{equation}
where the summation is over the gluon color and transverse polarization indexes which is not considered here. 
Consider an inclusive observable $O(N)$. 
To find the average value of $O(N)$ we have to calculate the matrix element $\langle O(N) \rangle$ over the final state of the scattering process.
We denote this final state as $\ket {P_{out}}$
\begin{equation}\label{eq_8}
\ket {P_{out}} = \Omega^\dag S \Omega \ket {P_0}
\end{equation}
The structure of the state $\ket {P_{out}}$ is a direct consequence of the statement that a creation of soft gluons can viewed as a quasielastic diffraction.
In a scattering process the coherence of the fast-projectile wave function is lost and slow gluons can be observed in a final state.
The role of the operator $\Omega^\dag$ in (\ref{eq_8}) is to suppress the coherent components in the state $S \Omega \ket {P_0}$.
So, the average $\langle O(N) \rangle$ is calculated as
\begin{equation}
\langle O(N) \rangle = \bra {P_{out}} O(N) \ket {P_{out}} = \bra {P_0} \Omega^\dag S \Omega O(N) \Omega^\dag S \Omega \ket {P_0}
\end{equation}
Also, it is useful to define the state $\ket {P_{in}}$ as 
\begin{equation}
\ket {P_{in}} = \ket {P_0}
\end{equation}
This allows to rewrite the elastic amplitude (\ref{eq_9}) in the very elegant form
\begin{equation}
S_{el} = \braket {P_{in}}  {P_{out}}
\end{equation}
In principle, it is possible to define new operator $S(Y)=\Omega^\dag S \Omega$ which can play the role of full scattering operator.
This operator produces the full set of inelastic final states. 
However, it is wrong to write  $\ket {P_{out}} = S(Y) \ket {P_0}$,
since here exists the intimate problem that $S(Y)$ strongly involves the Hilbert space of the target.
One should study $S(Y)$ in the full target-particle Hilbert space  $\ket {P_0} \otimes \ket {T_0}$

Now we are ready to define inclusive gluon spectrum $F_n$ in the final state as
\begin{equation} \label{eq_29}
F_n(x_1 \dots x_n) =\bra{P_{out}} N(x_1) \dots N(x_n) \ket{P_{out}}
\end{equation}
where rapidities $Y_1 \dots Y_n$ are omitted for brevity and $F_0=1$ is assumed.
Formally, quantities $F_n$ can be calculated from the more detailed theory by a direct computation. 
Note that usually the distributions $F_1$ and $F_2$ are in question, since they can be directly related to same experimental observables.

\section{The boost operator $\Omega$} \label{sec2}
From the perturbative expansion of the lightcone QCD Hamiltonian it is possible to calculate the operator $\Omega$. 
As it was shown is Ref. \cite{avp_Hqcd}, the first order gives the following Hamiltonian 
\begin{equation} \label{eq_10}
H=H_0+g\int\limits_{\vec x,x^-} \partial_i \tilde A_i^a \frac{1}{\partial_-}j_a^+
\end{equation}
where $j_a^+(\vec x,x^-)$ is the external color current. 
In the framework of high energy evolution, this current is associated with fast color sources, which are non-dynamical.   
The fast sources live in their own Hilbert space, in which the operator $j_a^+$ acts.
The Hilbert space of the slow modes is generated by the gluon creation operators $a^\dag$ and $a$, which form the quantum fields $\tilde A_i^a$ as
\begin{equation} \label{eq_16}
\tilde A_i^a(\vec{x},x^-)=\int\limits_{k^+>0}\frac{1}{\sqrt{2k^+}}\left(
         a^\dag_{k,a,i} e^{ik^+x^--i\vec{k}\vec{x}}+ 
         a_{k,a,i} e^{-ik^+x^-+i\vec{k}\vec{x}}
         \right) \frac{d^2k dk^+}{\sqrt{(2\pi)^3}}
\end{equation}
Before a rapidity boost, the phase space of slow gluons is zero, 
the Hamiltonian is simple $H_0$, and the projective state \ket {P_0} is an eigenstate of $H_0$.
After a boost, the slow gluons emerge and the state \ket {P_0} is not an eigenstate of $H$.
To find the modified projectile state \ket P, we can apply the standard stationary perturbation theory over the coupling constant $g$.
This way is preferable if a number of additional gluons is near 1. 
However if we need to handle arbitrary number of gluons, the most useful way is to construct a diagonalization operator $\Omega$, 
which also will be called as the boost operator.
The unitary operator $\Omega$ maps the bare projectile state $\ket {P_0}$, being an eigenstate of $H_0$, to the boosted projectile state $\ket P$, being an eigenstate of $H$.
Also, the transformation $\Omega H \Omega^\dag$ diagonalizes $H$ in the basis of the bare projectile states.

To find $\Omega$, let us express $H$ in terms of gluon operators $a^\dag$ and $a$, which are the field modes.
\begin{equation}
H_0 = \int\limits_k\frac{\vec k^2}{2k^+} a^\dag_{k,a,i} a_{k,a,i}
\end{equation}
\begin{equation} \label{eq_13}
H=H_0 + \int\limits_k \left({\bar c_{k,i,a}} a^\dag_{k,i,a} + c_{k,i,a} a_{k,i,a}\right)
\end{equation}
\begin{equation} \label{eq_14}
\bar c_{k,i,a}=\frac{g}{\sqrt{(2\pi)^3}} \int\limits_x  \frac{k^i\exp\left(ik^+x^--i\vec k \vec x\right)}{k^+\sqrt{2k^+}}  j^+_a(\vec x ,x^-)
\end{equation}
where we have evaluated 
\begin{equation}
\frac{1}{\partial_-}j^+_a(\vec x ,x^-)=\frac{1}{ik^+}j^+_a(\vec x ,x^-)
\end{equation}
since the transition current $j^+_a(\vec x ,x^-)$ is proportional to $e^{ik^+x^-}$, where $k^+$ is the transferred momentum.

We see that the interaction term in Eq. (\ref{eq_13}) is linear over the gluon creation operators.
Such dynamical systems is well known and the solution can be found in the theory of coherent states. 
Consider a simple harmonic oscillator with a linear interaction 
\begin{equation}
h=\omega a^\dag a + \bar c a^\dag+ca
\end{equation}
It can be diagonalized by the following unitary transformation  
\begin{equation}\label{eq_17}
\exp\left( \frac{-\bar ca^\dag+ca}{\omega}\right)
\end{equation} 
whose a normal ordered version is
\begin{equation} \label{eq_15}
\exp\left(-\frac{\bar c c}{2\omega^2}\right) \exp\left( -\frac{\bar ca^\dag}{\omega}\right) \exp\left( \frac{ca}{\omega}\right)
\end{equation}
To check the result (\ref{eq_17}), one can calculate the infinitesimal transformation of the free Hamiltonian $\omega a^\dag a$ 
for small $c$ as
\begin{equation}
\delta h=\left[\omega a^\dag a, \frac{-\bar ca^\dag+ca}{\omega}\right]=\bar c a^\dag +  ca
\end{equation}
Note, that if interaction part of a Hamiltonian is quadratic over the operators $a^\dag$ and $a$, 
then the diagonalization can be found in the framework of the Bogoliubov transformation, which is widely used in the condensed matter physics.

Returning to lightcone QCD, we see that the energy $\omega$ is replaced by $k^-=\vec k^2/2k^+$ which is the energy of a gluon mode.
Generalizing Eq. (\ref{eq_17}) to the field theory and using Eq. (\ref{eq_14}) we directly obtain
\begin{equation}
\Omega=\exp\left(\int\limits_k-\frac{1}{k^-} \bar c_{k,i,a}  a^\dag_{k,i,a} + \frac{1}{k^-}c_{k,i,a} a_{k,i,a}\right)
\end{equation}
\begin{equation}
\frac{\bar c_{k,i,a}}{k^-}=\frac{g}{\sqrt{(2\pi)^3}} \int\limits_x  \frac{2k_i\exp\left(ik^+x^--i\vec k \vec x\right)}{\vec k^2\sqrt{2k^+}}  j_a^+(x^-,\vec x)
\end{equation}
Performing the calculations, it is possible to obtain the following form of $\Omega$ 
\begin{equation} \label{eq_11}
\Omega=\exp \left( 2i\int\limits_x\tilde A_i^a(x^-,\vec x) b_i^a(x^-,\vec x)  \right)
\end{equation}
where $b_i^a$ obey the equations
\begin{equation}
\partial_i b_i^a=gj_a^+
\end{equation}
\begin{equation} \label{eq_18}
\partial_i b_j^a-\partial_j b_i^a=0
\end{equation}
The solution is
\begin{equation} \label{eq_23}
b_i^a(x^-,\vec x)=g\partial_i\frac{1}{\partial^2}j^+_a(x^-,\vec x)=\frac{g}{2\pi}\int \frac{x^i-y^i}{(\vec x-\vec y)^2} j^+_a(x^-,\vec x) d\vec y 
\end{equation}
Similarly to Eq. (\ref{eq_15}), the operator $\Omega$ can be expressed in a normal ordered form as
\begin{equation} \label{eq_12}
\Omega= \exp\left(-2\int\limits_{x,y} D_{ij}^{ab}(x-y) b^a_i(x)b^b_j(y)\right) :\Omega:
\end{equation}
where $x$ denotes the spatial vector $(x^-,\vec x)$ and $D_{ij}^{ab}$ is the equal-time soft gluon correlator
\begin{equation}
D_{ij}^{ab}(x-y)=\bra {0} A_i^a(x) A_j^b(y) \ket {0} 
=\delta_{ij}\delta^{ab} \delta(\vec x-\vec y)\int\limits_{k^+>0}\frac{1}{4\pi k^+}e^{-ik^+(x^--y^-)}
\end{equation}
Note that a special care should be used in the case of QCD, because the coherent coefficients $c$ and $\bar c$ do not commute
due to noncommutativity of the color charge density $\rho^a(\vec x)$.
The form (\ref{eq_12}) is very useful when one calculates matrix elements, 
since all gluon contractions are encapsulated into the first factor in Eq. (\ref{eq_12}) and the Wick's theorem is not needed. 

Also, in Eq. (\ref{eq_11}) for slow gluons and fast sources the integral $\int \exp(ik^+x^-) j^+_a dx^-$ is evaluated to $\rho^a(\vec x)$.
This fact is a consequence of the assumption that a matrix element between two projectile states  
\begin{equation} \label{eq_25}
\langle P_2 | \int \exp(ik^+x^-) j^+_a dx^- |P_1\rangle
\end{equation}
does not depend on $k^+$, since fast projectile partons is uniformly distributed above the cut-off over its longitudinal momentum  which is much larger than $k^+$.  
In other words, we does not measure a small change of longitudinal momentum of the projectile. 
This is some sort of an effective theory where some degrees of freedom is integrated out. 
So, the projectile partons has only transverse position $\vec x$ and the operator of charge density $\rho^a(\vec x)$ acts on the reduced Hilbert space. 
The lesson is that, working with an effective theory, we must always keep in mind a set of matrix elements which are considered in a particular case.
Using the assumption (\ref{eq_25}), we can rewrite the results of this Section as 
\begin{equation}
\Omega=\exp \left( 2i\int\limits_{\vec x}\tilde A_i^a(0,\vec x) b_i^a(\vec x)  \right)
\end{equation}
\begin{equation}
\partial_i b_i^a(\vec x)=\rho^a(\vec x)
\end{equation}
\begin{equation} \label{eq_26}
\Omega= \exp\left(-\frac{1}{2\pi}\int\limits_{\vec x,Y}  b^a_i(\vec x)b^a_i(\vec x)\right) :\Omega:
\end{equation}

Taking the rapidity interval $Y$ as an infinitesimal variable, 
matrix elements of an operator having $\Omega$ should be expanded only up to the linear terms over $Y$.
For example, JIMWLK equation can be obtained as
\begin{equation} \label{eq_20}
\frac{d\langle S \rangle}{dY}=\left. \frac{d}{dY} \langle \Omega^\dag S \Omega \rangle \right|_{Y=0}
\end{equation}
Inserting $\Omega$ from Eq. (\ref{eq_26}) into Eq. (\ref{eq_20}), we obtain all three terms of JIMWLK equation.

Formally, for finite $Y$ the formula (\ref{eq_11}) is not valid, because it does not include the self-interaction of the soft gluons,
which is usually leads to the quantum evolution. For inclusive processes, however, all terms are needed, 
since a gluon operator $O(N)$ reduces the power of $Y$ by an insertion of $\delta$-function into integrals.
And moreover, if we consider a process where gluons are emitted in a same rapidity, 
then the effects of the high energy evolution can be fully absorbed into the the projectile and target wave functions.
Gluon emissions with a rapidity gap we do not consider here. 

\section{Scattering is a diagonalization} \label{sec3}
Usually, the operator of quasielastic scattering $S$ in the high energy limit is calculated 
via the quasiclassical approximation of the path-integral formalism. 
The most known form is
\begin{equation} \label{eq_21}
S=e^{i\int\alpha_a(\vec x) \rho^a(\vec x)}
\end{equation}
where $\alpha_a(\vec x)$ is an external color field associated with a target state.
Eq. (\ref{eq_21}) says that
each color parton of the projectile at fixed transverse position is rotated by a corresponding element of the gauge group.
Obviously, this picture is explicitly asymmetric relatively a change of a target and a projectile. 
To make symmetric one, it is necessary to develop a quite complex formalism: the purely path integral \cite{Balitsky}, or the mixed type \cite{Kovner0810}.

Form the viewpoint of the lightcone Hamiltonian formalism the role of the scattering operator $S$ is not fully clarified yet.
To make the formalism consistent, we have to construct the symmetric operator $S$ purely within the Hamiltonian framework.
We show in this Section that the operator $S$ can be obtained from the boost operator $\Omega$ in the special limit.
This proves that $S$ can be directly derived from the lightcone Hamiltonian of QCD.

The starting point of the derivation is to represent the target and the projectile on the same footing in the Hilbert space.
Namely, consider a lightcone reference frame such that the target and the projectile are right-moving.
Both they form the two parton clusters localized in rapidity.
Boosting this system, we can view the incoming states only as an external color current that interact with slow gluon modes.
This setup allows to  apply the normal ordered version of the boost operator (\ref{eq_12}) where now the total current $j^+$ 
includes both the target and the projectile
\begin{equation} \label{eq_24}
j^+(x^-,\vec x)=j^+_P(x^-,\vec x)+j^+_T(x^-,\vec x)
\end{equation}
To extract the quasielastic channel of the scattering, we have to consider an outgoing wave function without additional gluons.
For the such matrix elements the operator $:\Omega:$ in Eq. (\ref{eq_12}) must be replaced by 1.
Thus, we have the key identity 
\begin{equation} \label{eq_22}
S(k^+)=\exp\left(-\frac{1}{2\pi k^+}  \int b^a_i(\vec x)b^a_i(\vec x)d\vec x\right)
\end{equation}
For quasielastic matrix elements the longitudinal momentum of the incoming states is not changed.
Hence, we must tend the momentum $k^+$ in Eq. (\ref{eq_22}) to zero. 
Note, that the limit $k^+\to 0$ pushs a soft gluon to off-shell.
Hence, we have to find an analytical continuation. 
Recall the rule
\begin{equation}
\frac{1}{x-i0}=i\pi\delta(x)+P\frac{1}{x}
\end{equation}
We need only the term $i\pi\delta(x)$, other terms leads to usual bremsstrahlung singularity of zero gluon $k^+$-momentum.
\begin{equation}
S=\exp\left(\frac{i}{2}\int  b^a_i(\vec x)b^a_i(\vec x)d\vec x\right)
\end{equation}
Using the solution (\ref{eq_23}) we have
\begin{equation}
S=\exp\left(\frac{ig^2}{2}\int  \rho^a(\vec x)\frac{1}{\partial^2}\rho^a(\vec x)d\vec x\right)
\end{equation}
Using Eq. (\ref{eq_24}) and selecting only the terms having both the target and the projectile, we finally obtain
\begin{equation} \label{eq_30}
S=\exp\left(ig^2\int  \rho^a_P(\vec x)\frac{1}{\partial^2}\rho^a_T(\vec x)d\vec x\right)
\end{equation}

We have seen that the operator $S$ is constructed from the charge densities $S[\rho_P, \rho_T]$.
Hence, its action on the boosted projectile in Eq. (\ref{eq_31}) can be written in a more detailed form.
Let $\tilde\rho_P$ be the operator of full color charge in a boosted projectile
\begin{equation}
\tilde\rho_P=\rho_P+\rho_A
\end{equation}
where $\rho_A$ is the operator of color charge in the Hilbert space of soft gluons
\begin{equation}
\rho_A^a(\vec x)=T^a_{8,bc} a^{\dag}_{b,i}(\vec x) a^{\phantom{\dag}}_{c,i}(\vec x)
\end{equation}
where $T^a_{8,bc}=if^{abc}$ is the group generator in the adjoint representation.

\section{AGK algebra} \label{sec4}
How to calculate easily an exclusive distribution? 
By definition it is just square of the gluons wave function $\ket {P_{out}}$ in the Fock space
and gives probability to find a given gluon configuration. 
If $n$-gluon wave function is $\Psi_n(x_1 \dots x_n)$ then the exclusive distribution $G_n$ is
\begin{equation}
G_n(x_1 \dots x_n)=\left | \Psi_n(x_1 \dots x_n)\right|^2
\end{equation}
The relation between $F_n$ and $G_n$ is well known from the very early studies \cite{Brown1972,Shei1972}.
However, the first recognition of AGK rules in this relation is given in Ref. \cite{avp_agk}. 
Namely, it can be shown that all relevant combinatorics of AGK rules are described by the relations in question.
The main theorem is
\begin{equation} \label{eq_1}
G[u]=F[u-1]
\end{equation}
where $F[u]$ and $G[u]$ is corresponding generating functional for $F_n$ and $G_n$ respectively
\begin{equation} \label{eq_2}
F[u]=\sum_{n=0}^\infty \frac{1}{n!}\int F_n(x_1 \ldots x_n) u(x_1) \ldots u(x_n) dx_1 \ldots dx_n
\end{equation}
\begin{equation}  \label{eq_3}
F_n(x_1 \ldots x_n)=\left. \frac{\delta^n F[u]}{\delta u(x_1) \ldots \delta u(x_n)} \right|_{u=0}
\end{equation}
and there are the equivalent relations for $G_n$.

Our task now is to express $G_n$ from known $F_n$. This will be called below as ``cutting rules''.
Using (\ref{eq_1}), (\ref{eq_2}), and (\ref{eq_2}) we directly obtain 
\begin{equation}
G_n=\sum_{k=n}^{\infty} G_n ^{(k)}
\end{equation}
\begin{equation} \label{eq_4}
G_n ^{(k)}=(-1)^{k-n}\frac{1}{(k-n)!}\int F_k(x_1\dots x_n,x_{n+1}\dots x_k) dx_{n+1}\dots dx_k
\end{equation}
To obtain a more familiar interpretation we consider topological cross sections.
The probability to find exactly $n$ gluons in the final state is
\begin{equation}
g_n=\frac{1}{n!}\int G_n(x_1 \dots x_n) dx_1 \dots dx_n
\end{equation}
Integration (\ref{eq_4}) we obtain
\begin{equation} \label{eq_5}
g_n^{(k)}=(-1)^{k-n} \frac{n!}{(k-n)! k!} f_k
\end{equation}
where $f_k$ is defined as 
\begin{equation}
f_k=\frac{1}{k!}\int F_k(x_1 \dots x_k) dx_1 \dots dx_k
\end{equation}
The relations (\ref{eq_5}) have a form of cutting rules and generate the standard AGK algebra of contributions to cross sections. 
Note that, since the relations (\ref{eq_5}) is quite formal, no new physic can be found here. 
Physical information is encapsulated in the inclusive coefficients $F_k$ and $f_k$.
The most optimistic conjecture, which is fulfilled in the toy models, is the total factorization for $x_1\dots x_k$ having same rapidity:
\begin{equation} \label{eq_6}
F_k=F_{projective}^k F_{target}^k F^k_{vertex}
\end{equation}
The factorization (\ref{eq_6}) holds in classical soft Reggeon Field Theory where the so called AGK cancellations take place.
This can be seen by t-channel cutting of arbitrary reggeon diagram and assigning the upper and lower parts to projectile and target respectively \cite{Kancheli}. 
In QCD, there is a problem that it is necessary to perform the averaging over the target fields, which break initial symmetry of the scattering.
Also note that here there are no any mentions about Pomerons neither soft nor hard. 
We think that there are no reasons to find any Regge-like quasiparticle is QCD, 
since a scattering in very complicated due to non-Abelian structure of the color gauge group $SU(3)$ \cite{avp_rz}.
Since there are no reggeon diagram technique in QCD, we shall work on formal algebraical level without a usage of any kind of Pomeron.

Let us relate the probabilities $g_k$ to cross sections of subprocesses.
Let $\sigma_k^{(n)}$ be the contribution to the total cross section from $k$ emitted gluons from inclusive distribution $F_n$ ($n$-order).
If $n>0$ then we obviously have
\begin{equation} \label{eq_7}
\sigma_n^{(k)}=g_n^{(k)}
\end{equation}
For $n=0$, there exits an additional step to separate unit from $S$-matrix. 
To find $\sigma_0^{(k)}$ we formally define an auxiliary quantity $\sigma_{tot}^{(k)}$ as 
\begin{equation}\label{eq_28}
\sigma_{tot}^{(k)}=\sigma_0^{(k)}+\sigma_1^{(k)}+\ldots\sigma_k^{(k)}=\sigma_0^{(k)}-g_0^{(k)}+\sum_{n=0}^k g_n^{(k)}
\end{equation} 
where we have used (\ref{eq_7}) and that $\sigma_n^{(k)}=0$ for $n>k$ due to (\ref{eq_4}).
Using (\ref{eq_5}) we can prove that 
\begin{equation}
\sum_{n=0}^k g_n^{(k)}=0
\end{equation}
Finally, from Eq. (\ref{eq_28}) we obtain
\begin{equation}
\sigma_0^{(k)}=g_0^{(k)}+\sigma^{(k)}_{tot}
\end{equation}
where $\sigma^{(k)}_{tot}$ should be calculated from an underlying theory.
In the toy models $\sigma^{(k)}_{tot}$ is just the contribution to the total cross section from the $n$-pomeron exchange.

Next, we define the cross section $\sigma_d$ as 
\begin{equation}
\sigma_0=\sigma_{el}+\sigma_d
\end{equation}
where $\sigma_d$ is the cross section of color diffraction, 
which is a process with a color exchange but without gluon emissions into a final state.
Usually, a such color diffraction gives the rapidity interval uniformly fulfilled by secondary hardrons from a color flux tube decay.
To construct the possible diffractive states we consider the projective as a set of color parton, 
even through the number of partons can vary. 
Then, each parton in the wave function is being arbitrary rotated in the color space.
In other words, the space of diffractive states is generated by the action of the gauge group on the projectile state.
One more relation, which should be mentioned here, is the optical theorem
\begin{equation}
\sigma_{tot}=2ReM
\end{equation}
where $M=1-S_{el}$ is the elastic amplitude.

Note that, in the toy models it is also possible to define the amplitudes $M^{(n)}$ that is exactly a contribution from the n-pomeron exchange.
Moreover, if the AGK cancellations take place, then $M^{(n)}$ and $F_n$ is directly related.
However, in our general case a relation between $M$ and $F_n$ is not assumed.

\section{Discussion} \label{sec5}

In this paper we have shown that in high energy QCD the relations (\ref{eq_29}), (\ref{eq_4}), 
and the boost operator $\Omega$ give the natural replacement of AGK rules.
In additional, in Eq. (\ref{eq_11}) we have derived $\Omega$ from ab initio QCD Hamiltonian in the first order. 
Also, to make the framework complete, in Eq. (\ref{eq_30}) we have obtained the operator of quasielastic scattering $S$ from the boost operator $\Omega$.
The results mean that we are able to calculate any quantity, both elastic and inelastic, both inclusive and exclusive.  
All these quantities can be directly extracted from the final state of scattering 
\begin{equation} \label{eq_32}
\ket {P_{out}}=\Omega^\dag S[\rho_P+\rho_A,\rho_T] \Omega \ket {P_0}
\end{equation}
where the dependence $\Omega[\rho_P, \tilde A]$ is assumed. 
The existence of a target state in Eq. (\ref{eq_32}) due to the operator $\rho_T$ leads to an additional averaging in the observables.

We have seen that the role of the original AGK theorem in still unclear in high energy QCD.
Instead of AGK, we have shown that QCD as a fundamental field theory allows to calculate directly both exclusive and inclusive distributions from the quantum lightcone Hamiltonian.
Only, the one half of the AGK relations emerges in QCD via the formal relation between these distributions.
This means that there are no practical reasons to reproduce original AGK in QCD, since the pomeron ideology is not easily applicable. 
The introduction of pomerons and RFT into QCD seems artificial.
There exists only one hope to find a subtle relation between an elastic amplitude and inclusive distributions. 
If so, this hypothetical relation will be a replacement for AGK cancellation in the QCD case.

The structure of the equation (\ref{eq_18}) can motivate us to hope that the full version of this equation is
\begin{equation} \label{eq_19}
F^a_{ij}[b]=\partial_i b_j^a-\partial_j b_i^a +g f_{abc} b^b_i b^c_j=0
\end{equation}
This indeed was noted in Ref. \cite{Kovner0501}, however, the first try to prove is given later in Ref. \cite{Kovner07}.
We have shown recently in Ref. \cite{avp11} that the problem in not fully clear, 
since the quantum Hamiltonian and canonical fields are not completely found in lightcone QCD. 
Both Refs. \cite{Hatta05} and \cite{Kovner07} use the special discontinues at $x^-=0$, which is unnatural from the viewpoint of general quantum theory. 
Currently, a rigorous prof of (\ref{eq_19}) in not known.

If we take $b_i$ obeying Eq. (\ref{eq_19}) and insert it into Eq. (\ref{eq_20}), then we shall obtain the so-called JIMWLK+ equation.
The same equation can be found in Ref. \cite{Hatta05} where the path integral technique is used.

There exist the question concerning physical integrability of an evolution equation.
In other words, does a solution of JIMWLK equation for finite $Y$ correspond to a finite-rapidity boost?
Mathematically, this question can be formulated by the checking of the following identity 
\begin{equation} \label{eq_27}
\Omega(Y_2)\Omega(Y_1)=\Omega(Y_2+Y_1)
\end{equation}
If so, any evolution equation can be integrated. 
Also, the identity (\ref{eq_27}) allows to calculate $\Omega(Y)$ recursively from the infinitesimal boost.
This procedure gives a ladder-type structure of $\Omega(Y)$, since any next step of the evolution is based on gluons from all previous steps.
Unfortunally, the boost operator from Eqs. (\ref{eq_26}) gives merely an infinitesimal boost and currently there are no tools to check the property (\ref{eq_27}).
In general, there can exist more complex terms in the Hamiltonian that create simultaneously two or more gluons in a wide rapidity range.  
In this case Eq. (\ref{eq_27}) can be violated.

\section*{Acknowledgments}
This work was supported by RFFI 11-02-01395-a grant. 
We thank N.V. Prikhod'ko for useful remarks and fruitful discussions.

\end{document}